\documentclass[a4j,10pt,onecolumn,oneside,notitlepage,final]{article}
\usepackage{alltt}
\usepackage{graphicx} 
\usepackage{amsmath}
\usepackage{amssymb}
\usepackage{amscd}
\usepackage{graphicx}

\title{Born's rule from statistical mechanics of classical fields: from hitting times to quantum probabilities}

\author{Andrei Khrennikov\\
International Center for Mathematical Modelling
\\in Physics and Cognitive Sciences\\
Linnaeus University,  V\"axj\"o, S-35195, Sweden\\
email: Andrei.Khrennikov@lnu.se}

\begin{document}
\maketitle

\begin{abstract} 
We show that quantum probabilities can be derived from statistical mechanics of classical fields.
We consider Brownian motion in the space of fields and show that such a random field interacting with threshold 
type detectors produces clicks at random moments of time.  And the corresponding probability distribution can be approximately described by the formalism of quantum mechanics. 
Hence, probabilities in quantum mechanics and classical statistical mechanics differ not so much as it is typically claimed.
The temporal structure of the ``prequantum random field'' (which is the $L_2$-valued Wiener process) plays the crucial role. Moments of detector's 
clicks are mathematically described as hitting times which are actively used in classical theory of stochastic processes. Born's rule appears as an approximate rule. In principle, the difference
between the ``precise detection probability rule'' derived in this paper and the conventional Born's rule can be tested experimentally.
In our model  the presence of the random gain in detectors playes a crucial role. We also 
stress the role of the  detection threshold. It is not merely a technicality, but the fundamental element of the model.
\end{abstract}

keywords: classical statistical mechanics, infinite-dimensional state space, random fields. quantum probability of detection, derivation of Born's rule, threshold detectors, asymptotics of error function, distribution of hitting times, Brownian motion in the space of fields

\section{Introduction}

In this paper we present a novel application of statistical mechanics with the infinite-dimensional state space, the space of 
classical fields. We study the problem of interaction of a random field, e.g., the electromagnetic field, with a detector 
of the thershold type. And we found that, for the natural random field, the Brownian motion in the space of fields (the $L_2$-valued Wiener 
process), statistics of clicks produced by such a detector can be {\it approximately} described by the Born's rule. In quantum mechanics (QM) 
the latter was postulated \cite{Born}. Whether it is possible to derive this rule from some natural physical principles is still the subject of intensive 
debates.\footnote{``The conclusion seems to be that no generally accepted derivation of the Born rule has been given to date, but this does not imply that such a derivation is impossible in principle.'' \cite{L}, cf.  \cite{CONT},  \cite{Hof}, \cite{Ojima4}.} Thus we show that classical
statistical mechanics of fields  (theory of random fields\cite{R1}--\cite{R3}) provides a possibility to derive the fundamental law of QM connecting theory 
with experimental statistics. 

We state again that the basic quantum rule is derived as an approximate rule. (Opposite to the conventional quantum mechanical viewpoint;
by the latter this rule is a ``precise rule''.)\footnote{We remark that recently 
the quantum optics group of G. Weihs performed a series of experiments testing the validity of the Born's rule in the 
triple-slit experiment \cite{Sorkin}. Whether the observed deviation from the Born's rule is of the fundamental nature or 
it is a consequence of  experimental technicalities such as e.g. nonlinearity of detectors is still an open question.}
The small parameter of the model (determining the magnitude of deviation of the Born's rule from the ``precise rule''
for the threshold type detectors interacting with the $L_2$-valued Wiener process) is the quantity 
\begin{equation}
\label{CL7}
\epsilon \equiv \frac{\overline{{\cal E}}_{\rm{pulse}}}{{\cal E}_d} <<1,
\end{equation}
where ${\cal E}_d$ is the detection threshold (having the physical dimension of energy) and $\overline{{\cal E}}_{\rm{pulse}}$
is the average energy of pulses  emitted (randomly) by  a source. Thus the theory is about random fields of low energy interacting 
with detectors with sufficiently high threshold. We derived the ``precise formula'' for 
the detection probability for the threshold type detectors. The basic idea behind this derivation is embedding
of the problem of detection moments for the threshold detection into well developed theory of {\it hitting times} \cite{HIT}.
(The latter is widely used in e.g. statistical radio-physics \cite{S3}.) This theory gives us the probability distribution 
of detection times. The final (precise) formula, see (\ref{CL6mn}), is quite complicated analytically. At the same time 
it is very simple from the viewpoint of numerical computation (summation of a series with quickly  decreasing terms). In principle, this 
formula can be tested experimentally.\footnote{The precise formula for the threshold detection probability contains the 
probability distribution of the detector's gain. Determination of this distribution is a complicated problem depending 
essentially on types of threshold detectors \cite{43}. We shall discuss this problem in section \ref{DETTR}.}

This paper can be considered as development of {\it prequantum classical statistical field theory} (PCSFT) \cite{KH1}--\cite{KH8}.
PCSFT provided representation of quantum averages and probabilities as averages and correlations with repspect to ensembles 
of classical fields; including entangled systems (cf.  Hofer\cite{Hof},de la Pena and Ceto \cite{DLP}, Casado et al. \cite{CSD}, Boyer \cite{BOY}, Cole \cite{CSD1}, Roychoudhuri \cite{Roychoudhuri}, Adenier\cite{Adenier}). So, QM and classical statistical mechanics became essentially closer than it was commonly believed. All quantum probabilistic quantities are represented in the classical probabilistic framework, cf. Manko et al. \cite{M1}--\cite{M9}, Hess et al. \cite{Hess}, \cite{Hess1}, De Raedt et al. \cite{Raedt0}, Garola et al. \cite{CLK}.
PCSFT reproduced these quantities  as 
averages of intensities of {\it continuous random fields.} However, quantum experimental statistics is based on {\it discrete events,} clicks of detectors.
To come closer to experiment, PCSFT should be completed by measurement theory describing the transition from continuous random intensities
to discrete random events of detection. 

The first model of the thershold detection (TSD) of continuous random fields serving PCSFT 
was proposed in \cite{A}. This model is  mathematically complicated. The ``prequantum random fields'' reproducing quantum statistics
are very singlular (of the white noise type). The state space of this model is the space of Schwartz tempered distributions. In \cite{KNN},
it was observed that the same result can be achieved by using essentially simpler model, namely, Brownian motion (Wiener process).
However, in  \cite{KNN} from the very beginning a rough asymptotics of the detection probabilities was used;  hence,  directly  the Born's rule 
(an approxiamte formula for detection probabilities) was derived.
In the present paper we use deeper results from theory of classical random fields (for hitting times) and  obtain the ``precise rule''
for threshold detection which can be compared with the Born's rule,  the approximate rule. Moreover, in \cite{KNN} only internal
degrees of freedom such as e.g. polarization were considered. In the present paper by starting with internal degrees of freedom we proceed to 
the really physical case of random fields on physical space.  (The latter model is classical statistical mechanics 
with infinite-dimensional state space.)

As was already stressed, our approach is based on embedding of the detection problem into theory of random hitting times. As a consequence, we shall use  the mathematical apparatus of classical probability and mathematical statistics. The error function of the Gaussian distribution and compllementary error function (its asymptotic expansion)  are our main mathematical tools, see, e.g., 
\cite{ERC0} for introduction and \cite{ERC} for applications.    

\section{Threshold detectors}

The detection procedure used in optical experiments is based on two steps threshold passing. The first threshold is so to say the ``hardware threshold''. This is a part of the solid state physical structure of the detector, we call it the {\it  fundamental threshold.} The second threshold
is in some sense the ``software threshold'' -- the {\it discrimination threshold.} This threshold is set by experiments, in the case of photo-multipliers and tungsten-based
superconducting transition-edge sensors (W-TESs), or directly in the process of fabrication, in the case of  silicon-avalanche-photodiodes \cite{MC}. In this paper we are basically interested in the discrimination threshold. However, we start with a brief discussion on fundamental thresholds. Hence, everywhere 
besides section \ref{FTR},{\it  the ``detection threshold'' is just the discrimination threshold.}

\subsection{Fundamental thresholds of detectors}
\label{FTR}

There is a difference in threshold's structure between {\it photo-multipliers}, as the one used in \cite{ASP}, \cite{Grangier}, \cite{Grangier1}, and more recent experiments (e.g.,   \cite{M1}--\cite{M5}) based on {\it silicon-avalanche-photodiodes.} However, detectors of both types are based on threshold processes. The total threshold 
is combined of a few different thresholds corresponding to different stages of the detection process. First, we describe 
{\it fundamental thresholds} corresponding to  photo-multipliers and silicon-avalanche-photodiodes. 
In photo-multipliers the fundamental threshold is given by the {\it work function,} i.e., ${\cal E}_{\rm{fund}}$ 
coincides with the work function.  In silicon-avalanche-photodiodes the fundamental threshold is given by the {\it band gap,}
i.e., ${\cal E}_{\rm{fund}}$  coincides with the band gap.

\subsection{Discrimination threshold} 
\label{DETTR}

A lot of noise is involved in the process of detection, e.g. \cite{MC}--\cite{POL2},
An important source of noise is the {\it multiplication process.}  For example, in photo-multipliers, once an electron has been extracted
from the metal, it is accelerated in vacuum by an electric field until its kinetic energy is enough to extract other
bounds electrons (secondary emission) when striking the surface of another metal surface (dynode), which will in
turn be used accelerated onto other dynodes to free more and more electrons, until that flow of electrons
becomes measurable as anodic current. The form and energy of output spikes vary significantly from one liberated energy 
carrier to another. Typically it is assumed that the gain is given by
\begin{equation}
\label{BE0c_L0}
 G=\alpha \xi^N,
\end{equation}
where $\xi$ is the multiplication factor of a single dynode, $\alpha$ is the fraction of photoelectrons collected by the multiplier
structure, and $N$ is the number of stages in the photomultiplier. In the most simple model, $\xi$ can be assumed to
follow a Poisson distribution about the average yield for each dynode, so that the gain is a compound Poisson process
over $N$ identical stages. However, experimental measurements of the single photoelectron pulse height spectra from
photo-multipliers exhibit a distribution with larger relative variance than predicted by the Poisson model (and in some
case with a decreasing exponential distribution instead of a peaked one), and there is thus no universal description
of multiplication statistics \cite{MC}.

Besides of noise produced by detectors, noise responsible for so called {\it dark counts} plays an important role. This noise of 
the random background is inescapable. In average spikes corresponding to signal detections differ in the amplitude from noise generated spikes.
This fact provides a possibility to filter  noise generated spikes by using a {\it discrimination threshold,} denote the later by the symbol
${\cal E}_{d}.$ The selection of this threshold is a delicate procedure. By selecting too low 
${\cal E}_{d}$ experimenter would count to many noise generated spikes, in particular, dark counts.  By selecting
it too high experimenter would discard too many spikes generated by the signal. In both cases quantum statistics would be essentially 
disturbed. 

During a last few years, a few leading groups in quantum optics started to use actively 
 with {\it Tungsten-based Superconducting Transition-Edge Sensors}   (W-TESs) -- the
ultra-sensitive microcalorimeters (in particular, in attempts to 
perform the EPR-Bell experiment with highly efficient detectors).  Suprisingly usage of such modern detectors
is also impossible without seting a proper discrimination threshold, see, e.g., \cite{POL1}, \cite{POL2}. for reviews. 
I guess that (since noise is everywhere) it is impossible to elaborate a photon detection procedure which is not of 
the threshold type.

\section{Threshold detection scheme}
\label{DD}

\subsection{Detection moment as hitting time}

We consider a threshold type detector with the threshold ${\cal E}_d.$ It interacts with a random field $\phi(s, \omega),$
where $s$ is time and $\omega$ is a chance parameter describing randomness. 
For a moment, we consider the ${\bf C}$-valued random field (complex stochastic process). In section \ref{DD1_X} we shall consider
random fields valued in finite-dimensional complex space $H.$
This correspond to detection of internal degrees of freedom such as e.g. polarization.   We stress that the real physical situation corresponds 
to random fields with infinite-dimensional state space, e.g., $H=L_2({\bf R}^3),$ the space of complex valued fields
$\phi:  {\bf R}^3 \to {\bf C}$ (or $\to {\bf C}^k),$ see section \ref{SPACE}

The energy of the field is given by 
${\cal E}(s, \omega) =\vert \phi(s, \omega)\vert^2$ (hence, the random field has the physical dimension $\sim \sqrt{\rm{energy}}).$ (In section 5 we shall also consider spatial variables. In this case the field has the physical dimension of the energy density, i.e., energy per volume.)  
A threshold detector clicks at the first moment of time $\tau(\omega)$ when signal's energy ${\cal E}$ multiplied by the gain $g,$ 
see section \ref{DETTR}, exeeds the threshold:
\begin{equation}
\label{CL}
g {\cal E}(\tau(\omega), \omega) \geq {\cal E}_d. 
\end{equation}
In the mathematical model the detection moment is defined as the first hitting time:
\begin{equation}
\label{CL1}
\tau(\omega) = \inf\{s\geq 0:  g {\cal E}(s, \omega) \geq {\cal E}_d\}.
\end{equation}

\subsection{PDF of the detection moment}

We proceed under the following basic assumption.  After arriving to  a threshold type detector a classical signal (random field)  behaves
inside this detector as the (complex) {\it Brownian motion,} i.e., the $\phi(s, \omega)$ is simply the {\it Wiener process}, the Gaussian process
having zero average at any moment of time
\begin{equation}
\label{CL1a}
E \phi(s, \omega) = 0.
\end{equation}
and the covariance function
\begin{equation}
\label{CL2}
E \phi(s_1, \omega) \overline{\phi(s_2, \omega)} = \min(s_1, s_2) \sigma^2;
\end{equation}
in particular, we can find average of its energy
\begin{equation}
\label{CL3}
E {\cal E} (s, \omega)= \sigma^2 s. 
\end{equation}
From this equation, we see that the coefficient $\sigma^2$ has the physical dimension of {\it power.}
 %This is the average power of the signal inteacting with a detector. At the moment of the ``signal's arrival''
%this power equals to zero  (in average, i.e.,  depending on pulses emitted by a source it fluctuates 
%slightly above zero). Then it increases 
We are interested in the probability distribution of the moments of the ${\cal E}_d$-threshold detection
for the energy of the Brownian motion. Since moments of detection are defined formally as hitting times, 
we can apply theory of hitting times \cite{HIT}, \cite{S1}.  Consider
\begin{equation}
\label{CL4}
\tau_a(\omega) =  \inf\{s\geq 0:  {\cal E}(s, \omega) \geq a^2\}=\inf\{s\geq 0:  \vert \phi(s, \omega)\vert \geq a\}.
\end{equation}
Its probability distribution function (PDF) is given by the complicated expression, see, e.g., Shyryaev \cite{S1}:
\begin{equation}
\label{CL5}
P(\tau_a \leq \Delta t) = 
4 \sum_{k=0}^\infty (-1)^k \Big[ 1 - \Phi\Big(\frac{a(1+2k)}{\sqrt{\sigma^2 \Delta t}}\Big)\Big],
\end{equation}
where $\Phi(x)= \frac{1}{\sqrt{2\pi}} \int_{-\infty}^x e^{-u^2/2} du$ is the PDF
of the standard Gaussian distribution. 
To show close relation to classical statistics, we also consider the {\it error function} \cite{ERC0}
$$
\rm{erf}(x) = \frac{2}{\sqrt{\pi}}\int_{0}^x e^{-u^2} du
$$
and the {\it complementary error function} \cite{ERC0}
$$
\rm{erfc}(x)  = 1-\rm{erf}(x)  = \frac{2}{\sqrt{\pi}} \int_x^{\infty} e^{-u^2} du .
$$
Now we select $a =\sqrt{\frac{{\cal E}_d}{g}}$ and set $\tau \equiv \tau_a.$ We obtain:
\begin{equation}
\label{CL6}
P(\tau(\omega)  \leq \Delta t) = 
2 \sum_{k=0}^\infty (-1)^k \rm{erfc}\Big((1+2k)\sqrt{\frac{{\cal E}_d}{2\sigma^2 \Delta t g}}\Big).
\end{equation}

\subsection{Asymptotics for the detection moment}

In coming considerations, $\Delta t$ is the {\it average duration of the interaction of a signal with a threshold detector.} 
The experimental scheme can be described in the following way. There is a source of random pulses of e.g. 
classical electromagnetic field. (Such pulses can be identified with wave packets used in the quantum formalism.) 
Each pulse propagates in space (we shall discuss its dynamics) and finally arrives to  a detector. The aforementioned temporal parameter 
$\Delta t$ is the (average) duration of interaction of an input pulse with a detector. 
The quantity 
\begin{equation}
\label{CL7_kk}
\overline{{\cal E}}_{\rm{pulse}}= \sigma^2 \Delta t
\end{equation}
 is the average energy of emitted pulses.\footnote{In fact, this is the average energy 
which is transmitted to  a detector by a pulse in the process of interaction. We proceed under the assumption 
that there are no losses and a detector ``eats'' (in average) the total energy 
of the emitted pulse.}
 We shall proceed under the basic assumption 
that this energy is essentially less than the threshold, i.e., 
\begin{equation}
\label{TT_YY}
\epsilon \equiv \frac{\overline{{\cal E}}_{\rm{pulse}}}{{\cal E}_d} <<1.
\end{equation}
This is a realistic assumption, since the discrimination threshold is set for the amplified signals from 
the detector and the gain producing this amplification is very large. Under this assumption the probability distribution $P(\tau(\omega)  \leq \Delta t)$ can be approximated by the first term in the series (\ref{CL7}):
\begin{equation}
\label{CL8}
P ( \tau( \omega )  \leq \Delta t) \approx 
2 \rm{erfc}(1/\sqrt{2 \epsilon g}).
\end{equation}
In further consideration we shall use the asymptotic behavior of the complementary error function\cite{ERC0}  for $x \to \infty:$
\begin{equation}
\label{CL6_a}
    \rm{erfc}(x) = \frac{e^{-x^2}}{x\sqrt{\pi}}\left [1+\sum_{n=1}^\infty (-1)^n \frac{1\cdot3\cdot5\cdots(2n-1)}{(2x^2)^n}\right ]=\frac{e^{-x^2}}{x\sqrt{\pi}}\sum_{n=0}^\infty (-1)^n \frac{(2n-1)!!}{(2x^2)^n}.
\end{equation}
To proceed towards derivation of the Born's rule, we are fine with the rough approximation  
\begin{equation}
\label{CL6_b}
    \rm{erfc}(x) \approx \frac{e^{-x^2}}{x\sqrt{\pi}}.
\end{equation}
(So, in our approach the Born's rule provides a rather rough approximation of detection probabilities.)
Hence, 
\begin{equation}
\label{CL8_a}
P ( \tau( \omega )  \leq \Delta t) \approx
2 \sqrt{\frac{2\epsilon g}{\pi}}  e^{-1/2\epsilon g}= 2 \sqrt{\frac{2\sigma^2 \Delta t g}{\pi {\cal E}_d}} exp\{-\frac{{\cal E}_d}{2\sigma^2 \Delta t g}\}.
\end{equation}

\subsection{Asymptotics in the presence of a random gain}

As was pointed in section \ref{DETTR}, the gain is by itself a random variable \cite{}.
 %We proceed under the assumation that the random signal emitted by a source and the gain in the detector are 
 %{\it independend random variables.} In this case randomness of the source and the gaim can be described by two different parameters,
 %$\phi=\phi(t, \omega_1)$ and $g=g(\oemga_2).$  Hence, dtection conditions (\ref{CL_pi}), (\ref{CL1_pi}) are generalized to 
%\begin{equation}
%\label{CL_pi}
%g(\omega_1)  {\cal E}(\tau(\omega_2), \omega_2) \geq {\cal E}_d. 
%\end{equation}
%In the mathematical model the detection moment is defined as the first hitting time:
Thus
\begin{equation}
\label{CL1_ll}
\tau(\omega)= \inf\{s\geq 0:  g(\omega) {\cal E}(s, \omega) \geq {\cal E}_d\}.
\end{equation}
Let us introduce  the random variable $\eta$ by setting
$$g= 1/\eta^2.$$ 
Thus in the presence of a random gain the detection moment  is given by 
$
\tau(\omega)) = \inf\{s \geq 0:   {\cal E}(s, \omega) \geq \eta^2 (\omega) {\cal E}_d\}.
$
Denote the probability distribution of $\eta$ by $P_\eta.$ We suppose that it has the density, 
denote the latter by the symbol $\rho_\eta(\lambda).$ 
To find the probability distribution of $\tau,$ we use {\it the formula of total probability:}
\begin{equation}
\label{CL1_aa}
P(\tau \leq \Delta t) =\int d \lambda  \rho_\eta(\lambda) P(\tau \leq \Delta t \vert \eta = \lambda),
\end{equation}
where the conditional probability has been already found:
\begin{equation}
\label{CL1_bb}
P(\tau \leq \Delta t \vert \eta = \lambda) \approx
2 \sqrt{\frac{2 \sigma^2 \Delta t}{\pi \lambda^2 {\cal E}_d}} exp\{-\frac{ \lambda^2{\cal E}_d}{2\sigma^2 \Delta t}\}. 
\end{equation}
Finally, we have 
\begin{equation}
\label{CL1_cc}
P(\tau \leq \Delta t) \approx
\frac{4 \sigma^2 \Delta t}{{\cal E}_d}
\int d \lambda  \frac{\rho_\eta(\lambda)}{\vert \lambda\vert } 
 \frac{exp\{-\frac{ \lambda^2{\cal E}_d}{2\sigma^2 \Delta t}\}}
{\sqrt{2\pi \frac{\sigma^2 \Delta t}{{\cal E}_d}}}.
\end{equation}

\subsection{The average number of clicks}

We now recall the meaning of the time interval $\Delta t:$ the average duration of the interaction of a signal with a detector.
Consider now a long run of experimental trials, emissions and detections, of the duration $T.$ Then the average number 
of clicks can be found as
\begin{equation}
\label{CL1_dd}
N \approx \frac{T}{\Delta t} P(\tau \leq \Delta t) \approx 
\frac{4 \sigma^2 T }{{\cal E}_d}
\int d \lambda  f_\eta(\lambda)  
 \frac{exp\{-\frac{ \lambda^2}{2\epsilon }\}}
{\sqrt{2\pi \epsilon }},
\end{equation}
where $f_\eta(\lambda)= \frac{\rho(\lambda)}{\vert \lambda\vert }.$ 
We now remark that $$
D_\epsilon(\lambda)= \frac{exp\{-\frac{ \lambda^2}{2\epsilon }\}}{\sqrt{2\pi \epsilon }}
$$ is the $\delta$-sequence for $\epsilon \to 0:
\lim_{\epsilon \to 0} D_\epsilon(\lambda)= \delta(\lambda).$ 
Hence, 
\begin{equation}
\label{CL1_dd_l}
N_\sigma \approx 
\frac{4 \sigma^2 T \langle \; \delta, f_\eta \rangle}{{\cal E}_d}.
\end{equation}

\section{Detection probability} 
\label{DD1}

\subsection{Derivation of the quantum  (Born's) rule for the detection probability}
\label{DD1_X}

Consider now a random signal $\phi(s, \omega)$  valued in the $m$-dimensional complex HIlbert space $H,$
where $m$ can be equal to infinity. 
Let $(e_j)$ be an orthonormal basis in $H.$ The vector-valued (classical)  signal $\phi(s, \omega)$ can be expanded 
with respect to this basis
\begin{equation}
\label{CL1_oy}
\phi(s, \omega)= \sum_j \phi_j(s, \omega) e_j, \; \phi_j(s, \omega)= \langle  e_j, \phi(s, \omega) \rangle.
\end{equation}
This mathematical operation is physically realized as spliting of the signal $\phi(s, \omega)$ into  components $\phi_j(s, \omega).$
These components can be processed through mutually disjoint channels, $j=1,2,...m.$
We now assume that there is a threshold detector in each channel, $D_1,..., D_m.$ We also assume that all detectors 
have the same threshold ${\cal E}_d>0$ and the same probability distribution of the gain, with the density 
$\rho_g(\lambda).$ 

Suppose now that $\phi(s, \omega)$ is the $H$-valued Brownian motion (the Wiener process in $H)$. 
This process is determined by the covariance
operator $B: H \to H.$ Any covariance operator 
is Hermitian, positive, and the trace-class  and vice versa. The complex Wiener process is characterized
by the Hermitian covariance operator. (We remark that complex-valued random signals are widely used in e.g. radio-physics.) 
We have, for $y\in H,$
$$
E \langle y, \phi(s, \omega) \rangle =0, 
$$ 
and, for $y_j\in H, j=1,2,$ 
$$
E \langle y_1, \phi(s_1, \omega)\rangle \langle \phi(s_2, \omega), y_2 \rangle =\min(s_1,s_2) \langle B y_1, y_2 \rangle.
$$ 
The latter is the covariance function of the stochastic process; in the operator form:
$
B(s_1, s_2) = \min(s_1,s_2) B.
$
We note that the dispersion of the $H$-valued Wiener process (at the instant of time $s)$ is given by 
\begin{equation}
\label{TR}
\Sigma_{s}^2\equiv  E \Vert \phi(s, \omega) \Vert^2= s \rm{Tr} B.
\end{equation} 
This is the average energy of this random signal at the instant of time $s.$ Hence,
the quantity
\begin{equation}
\label{TR_p}
\Sigma^2= \frac{\Sigma_{s}^2}{s}=  \rm{Tr} B
\end{equation} 
has the physical dimension of power; this is the average power of the signal of the Brownian motion type (it does not depend on time). 

\medskip

We also remark that by normalization of 
the covariance function for the fixed $s$ by the dispersion 
we obtain the operator,
\begin{equation}
\label{TRX2}
\rho= B/ \rm{Tr} B,
\end{equation} 
which formally has all properties of 
the {\it density operator} used in 
quantum theory to represent quantum states. Its matrix elements have the form
\begin{equation}
\label{TRX2j}
\rho_{ij} =b_{ij}/\Sigma^2.
\end{equation} 
These are dimensionless quantities. 
The relation (\ref{TRX2}) plays a fundamental role in our approach  \cite{KH1}--\cite{KH8}: {\it each classical random process generates  a quantum state
(in general mixed) which is given by the normalized covariance operator of the process.} One can proceed the other way around as well: 
each density operator determines a class of classical random processes \cite{Watanabe}.
\medskip

We now consider components of the random signal $\phi(s, \omega)$ and their correlations: 
\begin{equation}
\label{TR_p1}
E \phi_i(s_1, \omega) \overline{\phi_i(s_2, \omega)} = \min(s_1,s_2) \langle B e_i, e_j\rangle= \min(s_1,s_2) b_{ij}.
\end{equation} 
In particular, 
\begin{equation}
\label{TR_p1_a}
\sigma_j^2(s)\equiv E {\cal E}_j(s, \omega) \equiv E  \vert \phi_j(s, \omega)\vert^2 = s \; b_{jj}.
\end{equation} 
This is the average energy of the $j$th component at the instant of time $s.$ We also consider its average 
power:
\begin{equation}
\label{TR_p2} 
\sigma_j^2=b_{jj}. 
\end{equation} 
We remark that  the average power of the  total  random signal is equal to the sum of powers of its components: 
\begin{equation}
\label{TR_p3}
\Sigma^2= \sum_j \sigma_j^2.
\end{equation} 

Consider now a run of experiment of the duration $T.$ The average number of clicks for the $j$th detector can be approximately 
expressed as
\begin{equation}
\label{CL1_pp1}
N_j \equiv N_{\sigma_j}\approx 
\frac{ \sigma_j^2 T \; \langle \delta, f_\eta \rangle}{{\cal E}_d}. 
\end{equation}
This asymptotics is valid under the assumption (\ref{CL7_kk}). Here
$$
\overline{{\cal E}}_{\rm{pulse}}= \Sigma^2 \Delta t
$$
is the average energy of emitted pulses. (We state again that we consider the process of detection 
in the absence of losses, cf. footnote 3.)

The total number of clicks is (again approximately) given by 
\begin{equation}
\label{CL1_ppy}
N = \sum_j N_j \approx 
\frac{4 \Sigma^2 T \;  \langle \delta, f_\eta \rangle}{{\cal E}_d} . 
\end{equation}
Hence, for the detector $D_j,$  the probability of detection can be expressed as 
\begin{equation}
\label{CL1_ppo}
P_j= \frac{N_j}{N}\approx \frac{\sigma_j^2  }{\Sigma^2}= \rho_{jj}.
\end{equation}
This is, in fact, {\it the Born's rule for the quantum state $\rho$ and the projection operator} $\hat{C}_j= \vert e_j\rangle\langle e_j\vert$ 
on the vector $e_j:$
Hence, for the detector $D_j,$  the probability of detection can be expressed as 
\begin{equation}
\label{CL1_ppooo}
 P_j =\rm{Tr} \rho \hat{C}_j.
\end{equation}

\subsection{Ercf-rule for detection probability}
\label{SUBQ}
We state again that the expression (\ref{CL1_dd_l}) 
was based on an approximation of the PDF of the detection moment for the detection threshold ${\cal E}_d >> \overline{{\cal E}}_{\rm{pulse}}.$

By using the precise formula (\ref{CL6}) and by taking into account the presence of the random gain we obtain
$$
N_j = \frac{T}{\Delta t}  
\int d \lambda \rho_\eta(\lambda)  P(\tau_j \leq \Delta t \vert \eta =\lambda)
$$
\begin{equation}
\label{CL6m}
 =  \frac{2T}{\Delta t}
\sum_{k=0}^\infty (-1)^k \int d \lambda \rho_\eta(\lambda)  \rm{erfc}\Big((1+2k)\sqrt{\frac{\lambda^2 {\cal E}_d}{2 \sigma_j^2 \Delta t}}\Big).
\end{equation}
The total number of clicks in all detectors is the sum of $N_j:$
\begin{equation}
\label{CL6m_g}
N= \frac{2T}{\Delta t}
\sum_{k=0}^\infty (-1)^k \int d \lambda \rho_\eta(\lambda)  \sum_j \rm{erfc}\Big((1+2k)\sqrt{\frac{\lambda^2{\cal E}_d}{2 \sigma_j^2 \Delta t}}\Big).
\end{equation}
Hence, the probability of a click in the $j$th detector is given by sufficiently complex formula (generalized Born's rule):
\begin{equation}
\label{CL6mn}
 P_j =  \frac{
\sum_{k=0}^\infty (-1)^k \int d \lambda \rho_\eta(\lambda)  \rm{erfc}\Big((1+2k)\sqrt{\frac{\lambda^2{\cal E}_d}{2  \sigma_j^2 \Delta t}}\Big)}
{\sum_{k=0}^\infty (-1)^k \int d \lambda \rho_\eta(\lambda)  \sum_j \rm{erfc}\Big((1+2k)\sqrt{\frac{\lambda^2 {\cal E}_d}{2 \sigma_j^2 \Delta t}}\Big)}.
\end{equation}

We can easily rewrite this formula by using the original probablity distribution of the gain $g=g(\omega)$ and not of the random variable $\eta(\omega).$ 
(We state again that the latter random variable is related to the gain by the relation $g(\omega)= 1/ \eta^2(\omega).$)
We remark that densities of the probability distributions for $g$ and $\eta$ (for nonnegative $\eta)$ are connected by the rule:
\begin{equation}
\label{CL6m_h}
\rho_g(\lambda) = \frac{1}{2 \lambda^{3/2}} \rho_\eta\Big( \frac{1}{\sqrt{\lambda}} \Big), 
\; \; \rho_\eta(\lambda) = \frac{1}{2 \lambda^3} \rho_g\Big( \frac{1}{\lambda^2} \Big).
\end{equation}
We have
$$
N_j = \frac{T}{\Delta t}  
\int d \lambda \rho_g(\lambda)  P(\tau_j \leq \Delta t \vert g=\lambda)
$$
\begin{equation}
\label{CL6m_G}
 =  \frac{2T}{\Delta t}
\sum_{k=0}^\infty (-1)^k \int d \lambda \rho_g(\lambda)  \rm{erfc}\Big((1+2k)\sqrt{\frac{{\cal E}_d}{2 \lambda\sigma_j^2 \Delta t}}\Big).
\end{equation}
The total number of clicks in all detectors is the sum of $N_j:$
\begin{equation}
\label{CL6m_g_G}
N= \frac{2T}{\Delta t}
\sum_{k=0}^\infty (-1)^k \int d \lambda \rho_g(\lambda)  \sum_j \rm{erfc}\Big((
1+2k)\sqrt{\frac{{\cal E}_d}{2 \lambda\sigma_j^2 \Delta t}}\Big).
\end{equation}
Hence, the probability of a click in the $j$th detector is given by sufficiently complex formula (generalized Born's rule):
\begin{equation}
\label{CL6mn_G}
 P_j =  \frac{
\sum_{k=0}^\infty (-1)^k \int d \lambda \rho_g(\lambda)  \rm{erfc}\Big((1+2k)\sqrt{\frac{{\cal E}_d}{2 \lambda \sigma_j^2 \Delta t}}\Big)}
{\sum_{k=0}^\infty (-1)^k \int d \lambda \rho_g(\lambda)  \sum_j \rm{erfc}\Big((1+2k)\sqrt{\frac{{\cal E}_d}{2 \lambda \sigma_j^2 \Delta t}}\Big)}.
\end{equation}

As we have seen, it is not easy to derive the canonical Born's rule from (\ref{CL6mn}). Although this expression is very complex 
from the analytic viewpoint, it can be easy and quickly calculated by computer. It would be interesting to compare the result 
which can be obtained on the basis of (\ref{CL6mn}) with experimental data. TSD predicts that the experimental statistical data 
should match with this generalaized Born's rule better than with the canonical Born's rule. 

We also remark that, for sufficiently large values of ${\cal E}_d,$ the series inside expressions (\ref{CL6mn}) , (\ref{CL6mn_G}) coverge very quickly. Therefore by using only the first terms of these series we obtain 
very good approximation. Hence the following approximate  $\rm{Ercf}$-expressions\ for the detection probabilities can be used
\begin{equation}
\label{CL6mn7}
 P_j =  \frac{
\int d \lambda \rho_\eta(\lambda)  \rm{erfc}\Big(\sqrt{\frac{\lambda^2{\cal E}_d}{2  \sigma_j^2 \Delta t}}\Big)}
{\int d \lambda \rho_\eta(\lambda)  \sum_j \rm{erfc}\Big(\sqrt{\frac{\lambda^2 {\cal E}_d}{2 \sigma_j^2 \Delta t}}\Big)}.
\end{equation}
or directly in the gain variable
\begin{equation}
\label{CL6mn_G2}
 P_j \approx  \frac{
\int d \lambda \rho_g(\lambda)  \rm{erfc}\Big(\sqrt{\frac{{\cal E}_d}{2 \lambda \sigma_j^2 \Delta t}}\Big)}
{\int d \lambda \rho_g(\lambda)  \sum_j \rm{erfc}\Big(\sqrt{\frac{{\cal E}_d}{2 \lambda \sigma_j^2 \Delta t}}\Big)}.
\end{equation}
We remark that, although the gain variable has the straightforward physical meaning, and hence the representation (\ref{CL6mn_G2})  is better from the physical viewpoint, the $\eta$-representation is 
better from the mathematical viewpoint: the integrals in (\ref{CL6mn7}) are Gaussian. The same can be said about complete $\rm{Ercf}$-representations of the detection probabilities, cf.  
(\ref{CL6mn_G}) with (\ref{CL6mn}).

\section*{acknowledgements}

This paper was started during my visit   to Quantum Optics and Quantum Information Center of 
Austrian Academy of Science (March 2012); I would like to thank Anton Zeilinger and his coworkers,
especially Rupert Ursin and Sven Ramelow, for  the discussions on experimental verification of PCSFT and the knowledge transfer of the details on ``experimental technicalities''. 
This paper was finished during the school "Quantum Foundations and Open Systems", Joao Pessoa, Brazil (15-28 July, 2012) and I would like to 
thank Inacio de Almeida Pedrosa and Claudio Furtado for hospitality. I also thank my colleagues at the International Center for Mathematical Modeling (at LNU)  Irina Basieva, Astrid Hilbert, Borje Nilsson, and Sven Nordebo  for stimulating discussions. Irina Basieva performed numerical simulation on asymptotic convergence of the  detection probability to the quantum probability. The presence of the graphical illustrations essentially  imporved the presentation of the material of the paper.
This research was supported by the grant `` Mathematical Modeling'' of the 
Faculty of Natural Science and Engineering of Linnaeus University.

\end{document}